\documentclass[AMA,STIX2COL]{MRM}
\makeatletter

\DeclareRobustCommand{\greektext}{%
\fontencoding{LGR}\selectfont\def\encodingdefault{LGR}}
\DeclareRobustCommand{\textgreek}[1]{\leavevmode{\greektext #1}}
\ProvideTextCommand{\~}{LGR}[1]{\char126#1}
\usepackage{amsmath}
\usepackage[LGR,T1]{fontenc}
\usepackage{wasysym} 
\renewcommand{\thefigure}{\arabic{figure}}

\makeatother

\articletype{PREPRINT - Submitted to Magnetic Resonance in Medicine}%
\received{<day> <Month>, <year>}
\revised{<day> <Month>, <year>}
\accepted{<day> <Month>, <year>}
\begin{document}
\title{KomaMRI.jl: An Open-Source Framework for General MRI
Simulations with GPU Acceleration}

\author[1,2,3]{Carlos Castillo-Passi}{\orcid{https://orcid.org/0000-0001-6227-0477}}
\author[2,4]{Ronal Coronado}{\orcid{https://orcid.org/0000-0001-6735-2607}}
\author[5,6]{Gabriel Varela-Mattatall}{\orcid{https://orcid.org/0000-0001-6101-7218}}
\author[7]{Carlos Alberola-L\'{o}pez}{\orcid{https://orcid.org/0000-0003-3684-0055}}
\author[1,2]{Ren\'{e} Botnar}{\orcid{https://orcid.org/0000-0003-2811-2509}}
\author[2,3,4]{Pablo Irarrazaval}{\orcid{https://orcid.org/0000-0002-5186-2642}}

\address[1]{\orgdiv{School of Biomedical Engineering and Imaging Sciences}, \orgname{King's College London}, \orgaddress{\state{London}, \country{United Kingdom}}}
\address[2]{\orgdiv{Institute for Biological and Medical Engineering}, \orgname{Pontificia Universidad Católica de Chile}, \orgaddress{\state{Santiago}, \country{Chile}}}
\address[3]{\orgdiv{Millennium Institute for Intelligent Healthcare Engineering (iHEALTH)}, \orgname{Pontificia Universidad Católica de Chile}, \orgaddress{\state{Santiago}, \country{Chile}}}
\address[4]{\orgdiv{Electrical Engineering}, \orgname{Pontificia Universidad Católica de Chile}, \orgaddress{\state{Santiago}, \country{Chile}}}
\address[5]{\orgdiv{Centre for Functional and
Metabolic Mapping (CFMM), Robarts Research Institute}, \orgname{Western University}, \orgaddress{\state{London, Ontario}, \country{Canada}}}
\address[6]{\orgdiv{Department of Medical Biophysics, Schulich School of Medicine and Dentistry}, \orgname{Western University}, \orgaddress{\state{London, Ontario}, \country{Canada}}}
\address[7]{\orgdiv{Laboratorio de Procesado de Imagen}, \orgname{Universidad de Valladolid}, \orgaddress{\state{Valladolid}, \country{Spain}}}

\authormark{CASTILLO-PASSI \textsc{et al}}

\corres{Carlos Castillo-Passi, Institute for Biological and Medical Engineering, Pontificia Universidad Católica de Chile, Avda. Vicuña Mackenna 4860, Macul, Santiago, Chile.\\\email{cncastillo@uc.cl}}

\finfo{This work was supported by ANID - Millennium Science Initiative Program ICN2021\_004, and by Fondecyt 121074.\\\textbf{Word count:} 5252.}

\abstract{\section{Purpose}
To develop an open-source, high-performance, easy-to-use, extensible, cross-platform, and general MRI simulation framework (Koma).
\section{Methods}
Koma was developed using the Julia programming language. Like other MRI simulators, it solves the Bloch equations with CPU and GPU parallelization. The inputs are the scanner parameters, the phantom, and the pulse sequence that is Pulseq-compatible. The raw data is stored in the ISMRMRD format. For the reconstruction, MRIReco.jl is used. A graphical user interface utilizing web technologies was also designed. Two types of experiments were performed: one to compare the quality of the results and the execution speed, and the second to compare its usability. Finally, the use of Koma in quantitative imaging was demonstrated by simulating Magnetic Resonance Fingerprinting (MRF) acquisitions.
\section{Results}
Koma was compared to two well-known open-source MRI simulators, JEMRIS and MRiLab. Highly accurate results (with MAEs below 0.1\% compared to JEMRIS) and better GPU performance than MRiLab were demonstrated. In an experiment with students, Koma was proved to be easy to use, eight times faster on personal computers than JEMRIS, and 65\% of them recommended it. The potential for designing acquisition and reconstruction techniques was also shown through the simulation of MRF acquisitions, with conclusions that agree with the literature.
\section{Conclusions}
Koma's speed and flexibility have the potential to make simulations more accessible for education and research. Koma is expected to be used for designing and testing novel pulse sequences before implementing them in the scanner with Pulseq files, and for creating synthetic data to train machine learning models.
}

\jnlcitation{\cname{%
\author{Castillo-Passi C}, et al}.
\ctitle{KomaMRI.jl: An Open-Source Framework for General MRI
Simulations with GPU Acceleration}, \cjournal{Magn Reson Med}, \cvol{2023;XX:X--X}.}

\maketitle

\section{Introduction}

Numerical simulations are an important tool for analyzing and developing new acquisition and reconstruction methods in Magnetic Resonance Imaging (MRI). Simulations allow us to isolate and study phenomena by removing unwanted effects, such as hardware imperfections, off-resonance, and others. Additionally, with the increasing use of Machine Learning models, simulation becomes even more relevant, because it can be used to generate synthetic data for training,\cite{lundervold_overview_2019,maggiora_deepspio_2022} or to construct
signal dictionaries to infer quantitative measurements from
the acquired data.\cite{ma_magnetic_2013,kose_accurate_2020} Moreover, simulations are an excellent tool for education and training, as hands-on experience is a great way to assimilate the theoretical and practical components of MRI.\cite{treceno-fernandez_webbased_2019,treceno-fernandez_integration_2020,treceno-fernandez_magnetic_2021}

MRI simulators can be application-specific or general. Application-specific
simulators are efficient computationally and only consider a few relevant
effects (e.g. POSSUM,\cite{graham_realistic_2016,drobnjak_development_2006}
CAMINO,\cite{hall_convergence_2009} and others\cite{neher_fiberfox_2014,wissmann_mrxcat_2014,simpson_advanced_2017}).
A common simplification is that the acquired signal is equal to the
2D/3D Fourier transform of the image, not taking into account relaxation
during the acquisition, and other elements of MRI physics. On the
other hand, general simulators solve the Bloch equations making them
computationally intensive but usable in a wider range of applications.

\begin{figure*}[t]
\begin{centering}
\includegraphics[width=0.9\textwidth]{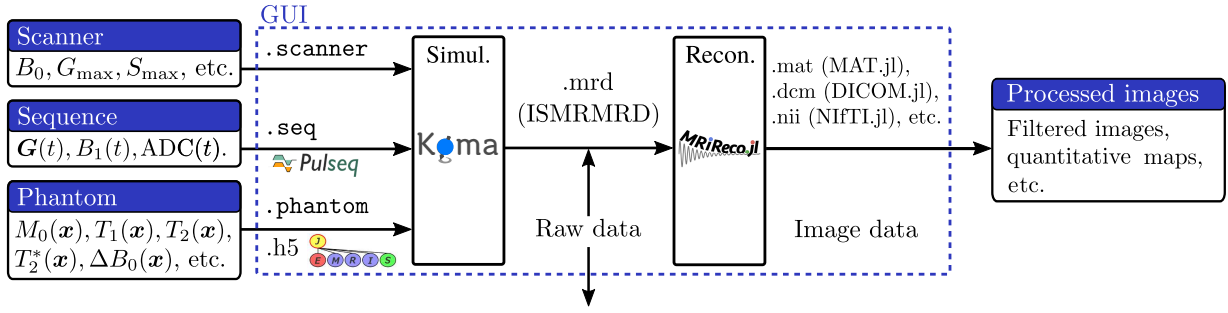}
\par\end{centering}
\caption{\enspace The simulation pipeline is divided in two steps: simulation, and reconstruction. Each arrow represents a file, and the pipeline
can be initiated from any stage of the data workflow. Note that we
are able to both read and write ISMRMRD files from the GUI.\label{fig:The-simulation-pipeline}}
\end{figure*}

Currently one of the most used general MRI simulators is JEMRIS.\cite{stocker_highperformance_2010}
This open-source simulator considers many properties of interest
in MRI, such as $M_{0}$, $T_{1}$, $T_{2}$, $T_{2}^{*}$, $\Delta B_{0}$,
movement, etc. JEMRIS only uses CPU multi-threading. Other alternatives are two closed-source simulators: MRISIMUL\cite{xanthis_mrisimul_2014}
and BlochSolver.\cite{kose_blochsolver_2017,kose_fast_2019} Both
are accelerated through the use of GPUs. Originally, MRISIMUL did not have a Graphical User Interface (GUI), but recently a cloud-based implementation called coreMRI has become available.\cite{xanthis_coremri_2019}
In 2017 a new open-source and GPU-accelerated simulator was introduced: MRiLab.\cite{liu_fast_2017} The main drawbacks of MRiLab are that it does not use self-contained sequence files
like JEMRIS, and that it was not designed to be extensible.

All of the previously mentioned simulators are written in C++ for
speed. In practice,  this may raise the bar for researchers in order to implement or modify these simulators.

The current open-source alternatives use MATLAB-based GUIs, resulting in non-intuitive interfaces. Furthermore, they do not support all Operating Systems (Table~\ref{tab:Overview-of-general-simul}).

\begin{table}
\caption{\enspace Overview of general MRI simulators.\label{tab:Overview-of-general-simul}}
\centering{}%
\resizebox{\linewidth}{!}{
\begin{tabular}{lcccc}
\toprule 
\textbf{Name} & \textbf{GUI} & \textbf{GPU} & \textbf{Open} & \textbf{Cross-platform}\tabularnewline
\midrule
JEMRIS\cite{stocker_highperformance_2010} & $\boldsymbol{\checkmark}$ & $\XBox$ & $\boldsymbol{\checkmark}$ & $\XBox$Windows\tabularnewline
MRISIMUL\cite{xanthis_mrisimul_2014} & $\XBox$ & $\boldsymbol{\checkmark}$ & $\XBox$ & $\XBox$\tabularnewline
BlochSolver\cite{kose_blochsolver_2017} & $\boldsymbol{\checkmark}$ & $\boldsymbol{\checkmark}$ & $\XBox$ & $\boldsymbol{\checkmark}$\tabularnewline
MRiLab\cite{liu_fast_2017} & $\boldsymbol{\checkmark}$ & $\boldsymbol{\checkmark}$ & $\boldsymbol{\checkmark}$ & $\XBox$Mac OS\tabularnewline
coreMRI\cite{xanthis_mrisimul_2014,xanthis_coremri_2019} & $\boldsymbol{\checkmark}$ & $\boldsymbol{\checkmark}$ & $\XBox$ & $\boldsymbol{\checkmark}$\tabularnewline
\bottomrule
\end{tabular}
}
\end{table}

We believe that an ideal simulator should be general, fast, easy to use,
extensible, open-source, and cross-platform. In this work, we developed
an MRI simulation framework built from the ground up to satisfy these
requirements.

To achieve these goals we made four important design decisions: programming language, compatibility with accepted standards, interface, and simulation method.

We chose the Julia programming language\cite{bezanson2017julia}
because its syntax is similar to MATLAB (widely used by the MRI community), its excellent GPU support,\cite{besard_effective_2019,besard_rapid_2019} and its speed is comparable to C/C++ (Julia is a compiled language).
This has been shown to be the case in other MRI applications such
as image reconstruction with MRIReco.jl,\cite{knopp_mrireco_2021}
where the authors achieved speeds on par with state-of-the-art toolboxes.\cite{uecker_martin_2021_4570601} In contrast to many other languages Julia can select the definition of a function to call at runtime via multiple dispatch. This is perhaps its most
powerful feature of Julia. This allowed us to use syntax that more closely follows mathematical notation.

The inputs to our simulation framework are the scanner parameters, the phantom, and the pulse sequence. For the latter we offer the possibility to program it directly in the code or alternatively to read it from a file in the standard Pulseq format.\cite{layton_pulseq_2017, opensource_, layton_open_} The output raw data is stored in the standard ISMRMRD format.\cite{inati_ismrm_2017} For reconstruction, our framework offers the possibility to use MRIReco.jl,\cite{knopp_mrireco_2021} any other reconstruction application that can read ISMRMRD, or direct programming of the code.

Using web technologies, we designed a GUI to improve accessibility for non-programmers and to facilitate the exploration of data and parameter tuning in a clear manner. This GUI also allows reading or writing the intermediate results.

We chose not to use an Ordinary Differential (ODE) solver, like DifferentialEquations.jl, but to handcraft an MRI-specific solver. This enabled us to use efficient solutions to the Bloch equations, and to implement an adaptive time-stepping based on information already available in the sequence. Both contributing to the simulation speed and accuracy.

We called our simulator ``Koma'', inspired by the Japanese word for spinning top, as its physics resemble MRI's.
\section{Methods}

In this section we start by describing in detail the simulation framework and its implementation (\ref{subsec:the-simulator}), we then describe the experiments we did for comparison (\ref{subsec:Experiments}), and finally, we showcase an application in quantitative imaging (\ref{subsec:MRF}).

\subsection{The Simulator \label{subsec:the-simulator}}

\subsubsection{Overview \label{subsec:Data-structures}}

KomaMRI.jl has three input objects that describe the scanner, the phantom, and the sequence
(Figure~\ref{fig:The-simulation-pipeline}):
\begin{itemize}
\item \texttt{Scanner}: Description of the hardware specifications such as $B_{0}$, maximum gradient $G_{\max}$ and slew rate $S_{\max}$. Other properties, such as inhomogeneity of the field $\Delta B_{0}\left(\boldsymbol{r}\right)$ and coil sensitivity should also be defined here (these are not available yet in the public version).
\item \texttt{Phantom}: Representation of the virtual object with properties such as the position $\boldsymbol{x}$ of the spins, proton density $M_{0}$, $T_{1}$, $T_{2}$, $T_{2}^{*}$, off-resonance $\Delta\omega$, non-rigid motion field $\boldsymbol{u}\left(\boldsymbol{x},t\right)$,
etc.
\item \texttt{Sequence}: Contains the Gradient waveforms $\boldsymbol{G}\left(t\right)$,
RF pulses $B_{1}\left(t\right)=B_{1,x}(t)+\mathrm{i}B_{1,y}(t)$ (where  $B_{1,x}(t)$ and $B_{1,y}(t)$ are the $x$ and $y$ components of the RF pulse), and data acquisition timing $\mathrm{ADC}\left(t\right)$.
\end{itemize}
We used the JLD2.jl package to implement our own HDF5-compatible file formats, \texttt{.scanner}, \texttt{.phantom} and \texttt{.seqk}, to load these objects. 
For the sequence, the simulator also accepts the newest versions of the Pulseq format (versions 1.2-1.4). This format gives our simulator more compatibility and convenience since it is also read by some real scanners.

The output of the simulator is written in the ISMRM Raw Data format\cite{inati_ismrm_2017}. This allows to test the reconstruction with different sources of data, simulated or real, and also allows to link the simulator with an external reconstruction.

\subsubsection{Physical and Mathematical Background\label{subsec:Simulator}}

Koma simulates the magnetization of each spin by solving the Bloch
equations in the rotating frame
\begin{align}
\frac{\mathrm{d}\boldsymbol{M}}{\mathrm{d}t} & =\gamma\boldsymbol{M}\times\boldsymbol{B}+\frac{\left(M_{0}-M_{z}\right)\hat{\boldsymbol{z}}}{T_{1}}-\frac{M_{x}\hat{\boldsymbol{x}}+M_{y}\hat{\boldsymbol{y}}}{T_{2}},\label{eq:Bloch}
\end{align}
with $\gamma$ the gyromagnetic ratio, $\boldsymbol{M}=\left[M_{x},M_{y},M_{z}\right]^{\mathrm{T}}$
the magnetization vector, and

\begin{align}
\boldsymbol{B}=\left[B_{1,x}\left(t\right),B_{1,y}\left(t\right),\boldsymbol{G}\left(t\right)\cdot\boldsymbol{x}\left(t\right)+\frac{\Delta\omega(t)}{\gamma}\right]^{\mathrm{T}}
 \label{eq: B definition}
\end{align}
 the effective magnetic field. $M_{0}$ is the
proton density, $T_{1}$ and $T_{2}$ are the relaxation times, and $\Delta\omega$ is the off-resonance, for each position.

The solution of Equation (\ref{eq:Bloch}) for a single spin is independent
of the state of the other spins in the system, a key feature that enables
parallelization.\cite{xanthis_mrisimul_2014}

\begin{figure}[h]
\begin{centering}
\includegraphics[width=0.35\textwidth]{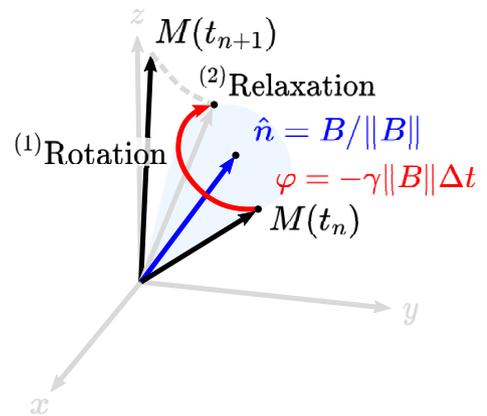}
\par\end{centering}
\caption{\enspace Solution of the Bloch equations for one time step can be described
by (1) a rotation and (2) a relaxation step. \label{fig:Operator splitting}}
\end{figure}

Our simulator uses a first-order splitting method\cite{graf_accuracy_2021}
to simplify the solution of Equation (\ref{eq:Bloch}). This reflects mathematically the intuition of separating the Bloch equations in a two-step process, rotation and relaxation, for each time step $\Delta t = t_{n+1} - t_{n}$ (Figure~\ref{fig:Operator splitting}). The rotation is described by
\begin{align}
\frac{\mathrm{d}\boldsymbol{M}^{(1)}}{\mathrm{d}t} & =\left[\begin{array}{ccc}
0 & \gamma B_{z} & -\gamma B_{y}\\
-\gamma B_{z} & 0 & \gamma B_{x}\\
\gamma B_{y} & -\gamma B_{x} & 0
\end{array}\right]\boldsymbol{M}^{(1)},\label{eq:rotation_Bloch}
\end{align}
with initial condition $\boldsymbol{M}^{(1)}(t_{n})=\boldsymbol{M}(t_{n})$, and the relaxation is described by
\begin{align}
\frac{\mathrm{d}\boldsymbol{M}^{(2)}}{\mathrm{d}t} & =\left[\begin{array}{ccc}
-\frac{1}{T_{2}} & 0 & 0\\
0 & -\frac{1}{T_{2}} & 0\\
0 & 0 & -\frac{1}{T_{1}}
\end{array}\right]\boldsymbol{M}^{(2)}+\left[\begin{array}{c}
0\\
0\\
\frac{M_{0}}{T_{1}}
\end{array}\right],\label{eq:relax_Bloch}
\end{align}
with $\boldsymbol{M}^{(2)}(t_{n})=\boldsymbol{M}^{(1)}(t_{n+1})$. Then, the magnetization at the end of the time step is $\boldsymbol{M}(t_{n+1})=\boldsymbol{M}^{(2)}(t_{n+1})$.

Furthermore, we define two regimes in the pulse sequence: excitation and precession. During the latter, the excitation fields are nulled: \textbf{$B_{x}=B_{y}=0$} in Equation (\ref{eq:rotation_Bloch}).
In the precession regime, the operator splitting method gives an exact solution, whereas during the excitation regime the method has $O\left(\Delta t^{3}\right)$ convergence.\cite{hazra_numerical_2018}

From this point forward, we will drop the vectorial notation for $\boldsymbol{M}$
and $\boldsymbol{B}_{1}$, and we will use $M_{xy}=M_{x}+\mathrm{i}M_{y}$
and $B_{1}=B_{1,x}+\mathrm{i}B_{1,y}$ to describe the simplifications
made in each regime.

The rotations during the excitation regime are stored in their spin-domain or SU(2) representation:

\begin{align}
\boldsymbol{Q}=\left[\begin{array}{cc}
\alpha & -\beta^{*}\\
\beta & \alpha^{*}
\end{array}\right],\quad\text{with }\left|\alpha\right|^{2}+\left|\beta\right|^{2}=1,
\end{align}
characterized by the Cayley-Klein complex parameters or Spinors for short $\left(\alpha,\beta\right)$.\cite{pauly_parameter_1991}
Spinors can represent any 3D rotation as

\begin{align}
\alpha & =\cos\left(\frac{\varphi}{2}\right)-\mathrm{i}\,n_{z}\sin\left(\frac{\varphi}{2}\right)\\
\beta & =-\mathrm{i}\,n_{xy}\sin\left(\frac{\varphi}{2}\right).
\end{align}

To solve Equation (\ref{eq:rotation_Bloch}) the parameters for the Spinors are $n_{xy}=B_{1}/\left\Vert \boldsymbol{B}\right\Vert $,
$n_{z}=B_{z}/\left\Vert \boldsymbol{B}\right\Vert $, and 
\begin{equation}
\varphi=-\gamma\left\Vert \boldsymbol{B}\right\Vert \Delta t\label{eq:angle_precession}
\end{equation}
is the phase accumulated due to $\boldsymbol{B}$, the effective magnetic field. Then, the application of a Spinor rotation to a magnetization element is described by the operation
\begin{equation}
\left[\begin{array}{c}
M_{xy}^{+}\\
M_{z}^{+}
\end{array}\right]=\left[\begin{array}{c}
2\alpha^{*}\beta M_{z}+\alpha^{*}{{}^2}M_{xy}-\beta{{}^2}M_{xy}^{*}\\
\left(\left|\alpha\right|{{}^2}-\left|\beta\right|{{}^2}\right)M_{z}-2\mathrm{Re}\left(\alpha\beta M_{xy}^{*}\right)
\end{array}\right].\label{eq:Rot_exc}
\end{equation}

For the precession regime, all the rotations are with respect to $z$,
and therefore they can be described with a complex exponential applied
to the transverse magnetization
\begin{equation}
M_{xy}^{+}=M_{xy}\mathrm{e}^{\mathrm{i}\varphi},\label{eq:Rotation_precession}
\end{equation}
where $\varphi$ is defined in Equation (\ref{eq:angle_precession}).

Finally, to solve the relaxation step described in Equation (\ref{eq:relax_Bloch})
the magnetization is updated by

\begin{align}
\left[\begin{array}{c}
M_{xy}^{+}\\
M_{z}^{+}
\end{array}\right]=\left[\begin{array}{c}
\mathrm{e}^{-\Delta t/T_{2}}M_{xy}\\
M_{z}\mathrm{e}^{-\Delta t/T_{1}}+M_{0}\left(1-\mathrm{e}^{-\Delta t/T_{1}}\right)
\end{array}\right].
\end{align}

The presented model solves the Bloch equations for a single isochromat, and by itself cannot simulate all the physical properties of interest in MRI. Other simulators implement more general equations like Bloch-Torrey,\cite{torrey_bloch_1956, stocker_highperformance_2010} to simulate diffusion, and Bloch-McConnell,\cite{mcconnell_reaction_2004, liu_fast_2017, asslander_generalized_2022} to simulate chemical exchange\cite{ward_new_2000, wu_overview_2016}, magnetization transfer\cite{wolff_magnetization_1989, henkelman_magnetization_2001a} and spin-lock effects.\cite{redfield_nuclear_1955, wang_1r_2015} 
 Besides non-rigid motion, we assume constant magnetic properties over time, so dynamic contrast-enhanced\cite{sujlana_review_2018} would require an extension of the model. 
 
 Nevertheless, incorporating other realistic effects like $T_{2}^{*}$ and diffusion could be easily added by increasing the number of spins. On the other hand, the Bloch-McConnell equations could also be implemented accurately with the operator splitting method.\cite{graf_accuracy_2021} While Koma is not as feature-complete as other well-established simulators, we focused on improving its speed, extensibility, and ease of use, and will keep adding more features in the future.

\begin{figure*}[t]
\begin{centering}
\includegraphics[width=1\textwidth]{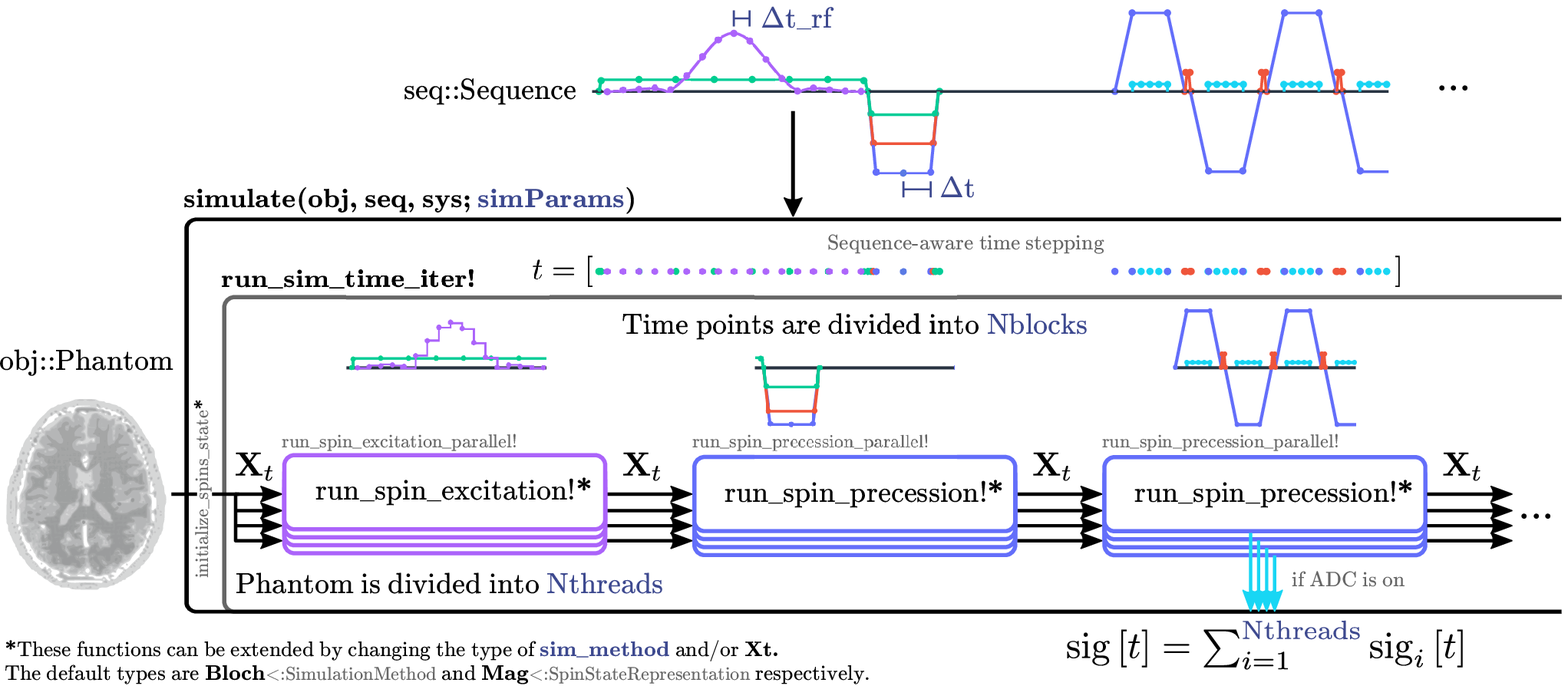}
\par\end{centering}
\caption{\enspace The sequence \texttt{seq} is discretized after calculating the required
time points in the wrapper function \texttt{simulate}. The time points
are then divided into \texttt{Nblocks} to reduce the amount of memory
used. The phantom \texttt{obj} is divided into
\texttt{Nthreads}, and Koma will use either \texttt{run\_spin\_excitation!}
or \texttt{run\_spin\_precession!} depending on the regime. If an ADC
object is present, the simulator will add the signal contributions
of each thread to construct the acquired signal $\mathrm{sig}\left[t\right].$
All the parameters: \texttt{Nthreads}, \texttt{Nblocks}, \texttt{\textgreek{D}t\_rf},
and \texttt{\textgreek{D}t}, are passed through a dictionary called
\texttt{simParams} as an optional parameter of the \texttt{simulate} function.\label{fig:GPU-CPU-parallel}}
\end{figure*}

\subsubsection{Simulation blocks, Regime Switching, and Sequence-Aware Time Stepping\label{subsec:Sequence-aware time stepping}}
To reduce the memory usage of our simulator, we subdivided time into \texttt{Nblocks} (Figure~\ref{fig:GPU-CPU-parallel}). Koma classifies each block in either the excitation regime or the precession regime before the simulation.

For precession blocks, we can improve the accuracy of the simulations by using the integral representation of Equation (\ref{eq:Rotation_precession}), obtained by applying the limit as $\Delta t\rightarrow0$ of iterated applications of Equation (\ref{eq:Rotation_precession}), giving a phase of

\begin{align}
 \varphi &=-\gamma \int_{t_i}^{t_{i+1}}B_z(\tau)\,\mathrm{d}\tau \\&=-\gamma\int_{t_i}^{t_{i+1}}\left(\boldsymbol{G}\left(\tau\right)\cdot\boldsymbol{x}\left(\tau\right)+\frac{\Delta\omega(\tau)}{\gamma}\right)\,\mathrm{d}\tau.
\end{align}

Assuming that during the $i$-th simulation block ($t\in\left[t_i,t_{i+1}\right]$) the gradients $\boldsymbol{G}\left(t\right)$ are piece-wise linear functions, and $\boldsymbol{x}\left(t\right)$ and $\Delta\omega(t)$ are approximately constant, then, if we use the trapezoidal rule to obtain the value of this integral, we will obtain an exact result by sampling just the vertices of $\boldsymbol{G}\left(t\right)$, greatly reducing the number of points required by the simulation. We will only need intermediate points in the case of motion and for recording the sampling points as required by the Analog to Digital Converter (ADC). The user can control the time between intermediate gradient samples with the parameter $\Delta$\texttt{t} (Figure~\ref{fig:GPU-CPU-parallel}).

We can do something similar with $B_{1}\left(t\right)$ in the excitation
regime. If we assume $B_{1}\left(t\right)$ is a piece-wise constant
function (or concatenation of hard pulses), then Equation (\ref{eq:Rot_exc}) will give an exact solution to Equation (\ref{eq:rotation_Bloch}).\cite{bittoun_computer_1984} The parameter $\Delta\texttt{t\_rf}$ manages the time between RF samples (Figure~\ref{fig:GPU-CPU-parallel}).

Thus, Koma uses the rationale mentioned above to: (1) call different methods based on the regime of each block, while also (2) obtaining a variable time stepping schedule that adapts to the sequence needs. We named the latter sequence-aware time stepping (Figure~\ref{fig:GPU-CPU-parallel}). While this concept is not new per se,\cite{xanthis_mrisimul_2014, liu_fast_2017, puiseux_numerical_2021} in Koma we directly calculate the required simulation points from Pulseq files or a designed \texttt{Sequence}  and then provide a convenient \texttt{DiscreteSequence} type to the user to simulate with our default \texttt{Bloch<:SimulationMethod} or to use in their custom simulation method. We comment further into this in subsection~\ref{subsec:Extensibility}.

\subsubsection{GPU/CPU Parallelization\label{subsec:GPU/CPU-parallelization}}

One key advantage of using Julia is its support for CPU parallelization using macros like \texttt{Threads.@threads} before a \texttt{for} loop, or the package \texttt{ThreadsX.jl}. Using these resources, we increased the simulation speed by separating the Bloch
calculations into \texttt{Nthreads}. This separation is possible as all magnetization vectors are independent of one another. To ensure thread safety, we stored the acquired signals per thread in different matrices to add them later into a signal matrix $\mathrm{sig}[t]$ (Figure~\ref{fig:GPU-CPU-parallel}). 

Julia also has native GPU support using the package \texttt{CUDA.jl}. This package supports operations using \texttt{CuArray} types which run as GPU kernels, but direct GPU kernel programming is also possible. To transfer variables between CPU and GPU memory, we used the packages \texttt{Adapt.jl} and \texttt{Functors.jl}. These packages let us transparently transfer our data types from CPU to GPU without losing the type abstractions. Then, the transfer looks like \texttt{obj = obj |> gpu}. Our data types, \texttt{Phantom}, \texttt{DiscreteSequence}, and \texttt{Mag<:SpinStateRepresentation} were used in this way, to then perform the simulation inside the functions \texttt{run\_spin\_excitation!} and \texttt{run\_spin\_precession!}.

It was important to ensure the type stability of our simulation functions to give enough information to the compiler to infer the concrete type of every variable, enabling high performance. Moreover, we had special care to perform in-place operations and not generate unnecessary variable copies using the \texttt{@view} macro in the functions \texttt{run\_spin\_excitation\_parallel!}, \texttt{run\_spin\_precession\_parallel!}, and \texttt{run\_sim\_iter!}. Finally, we used NVIDIA Nsight Systems to profile GPU performance with the \texttt{NVTX.@range} and \texttt{CUDA.@profile} macros.

\subsubsection{Reconstruction\label{subsec:Reconstruction}}

For the image reconstruction, we used MRIReco.jl,\cite{knopp_mrireco_2021}
a reconstruction framework written in Julia with comparable performance
to BART,\cite{uecker_martin_2021_4570601} a state-of-the-art reconstructor.
Coincidentally, as of version 2.9, JEMRIS uses BART as reconstructor.
The obtained image can be saved in multiple formats such as \texttt{.mat},
\texttt{.dcm}, and others, as shown in Figure~\ref{fig:The-simulation-pipeline}.

The reconstructor can also load directly the raw data from an ISMRMRD file, skipping the simulation.

\begin{figure*}
\begin{centering}
\includegraphics[width=\textwidth]{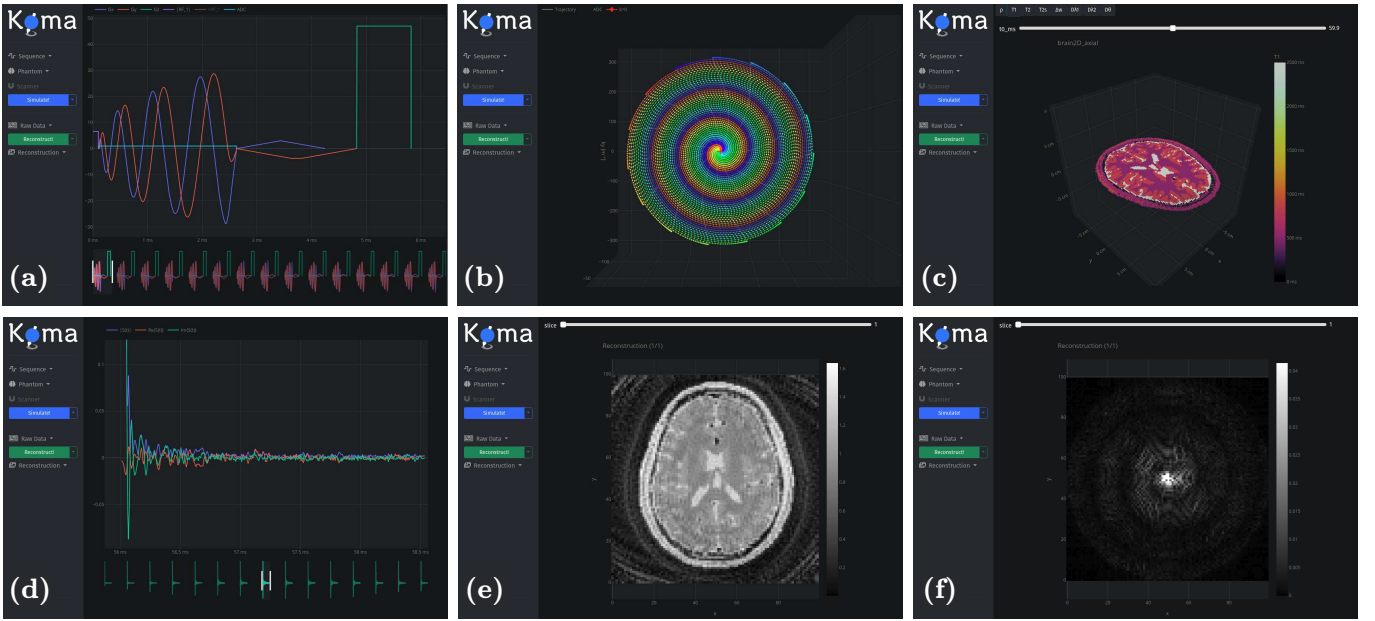}
\par\end{centering}
\caption{\enspace 
Koma's GUI: \textbf{(a)} Sequence (interleaved spiral), \textbf{(b)} $k$-space, \textbf{(c)} Phantom, \textbf{(d)} Raw Data, \textbf{(e)} Image, and \textbf{(f)} $k$-space data (Fourier transform of \textbf{(e)}). The menu at the left of the GUI tries to mimic the pipeline of Figure~\ref{fig:The-simulation-pipeline}, where the blue button calls Koma and the green button calls MRIReco.jl. \label{fig:GUI}}
\end{figure*}

\subsubsection{Graphical User Interface\label{subsec:Graphical-User-Interface}}

For the GUI we used Blink.jl, a framework to develop applications using web technologies. This package is a wrapper of Electron, and can serve HTML content in a local window.
The communication between Julia and this web page is done using JavaScript.

The GUI allows the user to easily plot the sequence, $k$-space, phantom, acquired signal, and reconstructed images (Figure~\ref{fig:GUI}). Plots are done using the PlotlyJS.jl package, which also allows to export them to \texttt{.svg} files.

\subsubsection{Extensibility\label{subsec:Extensibility}}

As we mentioned in the Introduction, in Julia, functions use different methods based on the input types via multiple dispatch. We used this to specialize the simulation functions for a given \texttt{sim\_method<:SimulationMethod} specified in \texttt{simParams}. For a given simulation method, the function \texttt{initialize\_spin\_state} outputs a variable \texttt{Xt<:SpinStateRepresentation} that is passed through the simulation (Figure~\ref{fig:GPU-CPU-parallel}). For the default simulation method \texttt{Bloch}, the spin state is of type \texttt{Mag}, but can be extended to a custom representation, like for example EPGs\cite{weigel_extended_2015} or others. Then, the functions \texttt{run\_spin\_excitation!} and \texttt{run\_spin\_precession!} can be described externally for custom types \texttt{sim\_method} and \texttt{Xt}, extending Koma's functionalities without the need of modifying the source code and taking advantage of all of Koma's features.

\subsection{Experiments \label{subsec:Experiments}}

To test Koma, we performed two kinds of experiments: one to compare the quality of the results and the execution speed, and the second one to compare its usability.

\subsubsection{Simulation Accuracy and Speed\label{subsec:Comparison-with-JEMRIS-Methods}}

To test the accuracy of our simulator, we compared Koma
with the latest version of JEMRIS (v2.9), which has been compared with real MRI acquisitions.\cite{veldmann_opensource_2022}.

We did 2D experiments with
different number of spins to look at the scalability of the simulations.
The experiments were as follows:
\begin{enumerate}
\item[(a)] EPI acquisition of a column of spins (1D data in a 2D image). This column was subdivided into four segments of length $l=50\,\mathrm{mm}$. The properties of each segment were $M_0=[1, 0.5, 1, 0.5]$, and $T_1=T_2=[100, 50, 100, 50] \,\mathrm{ms}$.

\item[(b)] EPI acquisition of two concentric circles ($R=50\,\mathrm{mm}$ and $r=25\,\mathrm{mm}$) with a constant frequency offset for the circle in the middle ($\Delta \omega = 200\,\mathrm{rad/s}$). Both circles had the same $M_0=1$ but different relaxation times ($T_1^{r}=T_2^{r}=50\,\mathrm{ms}$, and $T_1^{R}=T_2^{R}=100\,\mathrm{ms}$).

\item[(c)] EPI acquisition of a human brain, with properties obtained from the BrainWeb database,\cite{aubert-broche_new_2006} including a realistic off-resonance field ($\Delta\omega$'s ranging from $-400$ to $1200\,\mathrm{rad/s}$).

\item[(d)] EPI acquisition of the same brain of (c) without off-resonance but with motion. We applied a displacement field in the $y$-direction of $u_y(\boldsymbol{x},t)=v_y t$ with $v_y=0.1\,\mathrm{m/s}$.

\item[(e)] Spiral acquisition of the same brain of (c) without off-resonance.
\end{enumerate}

The EPI was a single-shot sequence with $\mathrm{TE}=100\,\mathrm{ms}$. The spiral acquisition was a single-shot sequence with $\mathrm{TE}=0.1\,\mathrm{ms}$. Both were Gradient Echo sequences with the same $\mathrm{FOV}=230\times230\,\mathrm{mm}$ and spatial resolution/voxel size $\Delta x = 2.3\,\mathrm{mm}$, while the resolution of the phantoms were $\Delta x_{\mathrm{obj}}=1\,\mathrm{mm}$. All the images were reconstructed in a $100\times100$ matrix with $\mathrm{FOV}=230\times230\,\mathrm{mm}$. Both sequences used hard RF pulses.

For the reconstruction of the spiral data for both simulators we used MRIReco.jl, and not BART for JEMRIS since it uses a different implementation of
the NUFFT algorithm, and we wanted to keep the image comparison as fair as possible.

To compare the simulation accuracy, for each experiment the signals were normalized by JEMRIS' signal maximum, and then we calculated the Mean Absolute Error (MAE) between them, $\mathrm{MAE}\left(\boldsymbol{x},\hat{\boldsymbol{x}}\right)=\frac{1}{n}\sum_{i=1}^{n}\left|x_{i}-\hat{x}_{i}\right|$. The errors are shown as a percentage of JEMRIS' signal maximum ($k$-space center).

All these examples were run in a computer with an 11th Gen Intel Core
i7-1165G7 @ 2.80GHz$\times$8, with four physical cores, 16 GB
RAM, a GPU GTX 1650 Ti (4 GiB of memory), and an eGPU RTX 2080 Ti (11 GiB of memory). For these examples, we only reported times for the faster GPU RTX 2080 Ti.

On the other hand, we compared the speed of our simulations against MRiLab, an open-source GPU-accelerated MRI simulator. For this, we replicated MRiLab's gradient echo multi-shot spiral sequence "PSD\_GRE3DSpiral" ($\mathrm{TE}=50\,\mathrm{ms}$, $\mathrm{TR}=10\,\mathrm{s}$, and $\Delta x = 2.5 \,\mathrm{mm}$), which contains a slice-selective sinc RF pulse with a slice thickness of $6\,\mathrm{mm}$, in conjunction with an 8-shot spiral acquisition. We selected this sequence to stress test both simulators, as it has arbitrary waveforms for both RF and gradients pulses. We used the standard resolution ($\Delta x_{\mathrm{obj}}=2\,\mathrm{mm}$) 3D brain phantom present in MRiLab. We followed their simulation procedure and only simulated in a slab of $N_{\mathrm{spins}}=20,630$ contained within the slice selection. We paid special attention to matching the number of time points and the spiral waveforms\cite{glover_simple_1999} to get comparable results. We ran this simulation in the CPU, and both GPUs, for both simulators 20 times for each device and calculated the mean and standard deviation.

We used the same computer as in the accuracy experiments, but we ran the test with both GPUs. We also simulated a similar sequence in JEMRIS to have as a reference.

\begin{figure*}
\centering{}\includegraphics[width=\textwidth]{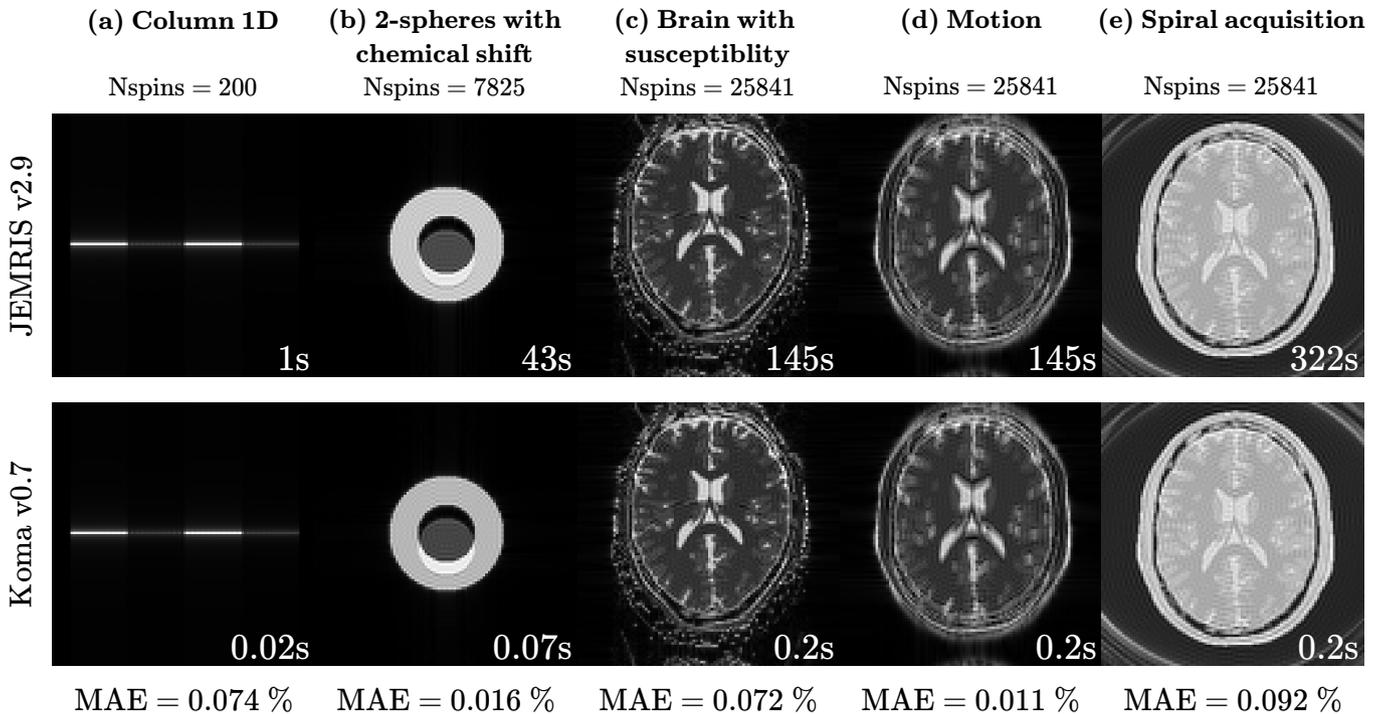}\caption{\enspace Simulations\label{tab:cores_time} \textbf{(a)}, \textbf{(b)}, \textbf{(c)}, and \textbf{(d)} used an EPI acquisition
with $\mathrm{TE}=100\,\mathrm{ms}$, but \textbf{(e)} used a
spiral acquisition with $\mathrm{TE}=0.1\,\mathrm{ms}$. All the simulations
were reconstructed in a matrix of $100\times100$ with $\mathrm{FOV}=230\times230\,\mathrm{mm}$. To compensate for
the differences in the NUFFT reconstruction between BART and MRIReco.jl,
both results in \textbf{(e)} used MRIReco.jl. We compared the accuracy of our simulations against JEMRIS by calculating the Mean Absolute Error (MAE) of the normalized simulated signals.}\label{fig:Simulations-(a),-(b),}
\end{figure*}

\subsubsection{User Experience}

To compare the ease of use for first-time users, we designed a pilot experience with students of an Imaging course in Engineering, where they learned some fundamentals of MRI. The
experience consisted in identifying the artifacts generated by the
presence of different degrees of off-resonance and motion (like examples (c) and (d) of Figure~\ref{fig:Simulations-(a),-(b),}). They
were to compare the artifacts of Gradient Echo and Spin Echo EPI acquisitions. 

There were 19 students in the class. We divided them in two groups. The first half of the students programmed the Gradient Echo sequence and the other half the Spin Echo sequence. Each student performed 6 experiments per simulator with brain phantoms ($N_\mathrm{spins}=5,890$) with different levels of off-resonance and motion. 

For the first part of the assignment, which tested from the installation to the first simulation, they used JEMRIS v2.8.3, MRiLab 1.3, and Koma v0.3.8. For the second part, they used JEMRIS and Koma to generate the EPI sequences
and different phantoms separately. We made tutorials to help them install the simulators and gave them examples of how to set up a simulation in all software packages.
They used the same phantoms and sequences. For the sequence, they had to calculate the timings and gradients strengths with the information taught during the course.

To gather information about how long it took them to perform
each task and their perceived level of difficulty, they filled out a form (available in Koma's GitHub) and returned it with their reports.
The difficulty
level of each task was rated on a Likert scale from 1 to 5, with 1
being hard and 5 being easy.

The students ran the simulations on their personal computers.

\subsection{Magnetic Resonance Fingerprinting \label{subsec:MRF}}

We used our simulator to showcase its potential, simulating a quantitative MRF acquisition.\cite{ma_magnetic_2013}

The sequence was a radial bSSFP sequence ($N_{\mathrm{spokes}}=158$) to acquire
the signal fingerprints for each pixel. The MRF sequence had an initial inversion pulse with TI of $50\,\mathrm{ms}$. For the first 500 TRs of the sequence, we used a Perlin-noise\cite{perlin_image_1985} flip angle pattern, and for the last 500 TRs, we used a noisy sinusoid flip angle pattern between $0\,\mathrm{deg}$ and $80\,\mathrm{deg}$ similar to Ma.\cite{ma_magnetic_2013} The TRs were randomly distributed between 14.5 and 18.0 ms, and a constant TE of 5 ms was used. A dictionary was generated
to do the fingerprint matching with the following ranges of $T_{1}$
and $T_{2}$: $T_{1}$ ($300$ to $2,500\,\mathrm{ms}$, every $10\,\mathrm{ms}$)
and $T_{2}$ ($40$ to $350\:\mathrm{ms}$, every $4\,\mathrm{ms}$). 

The \texttt{Phantom} object was a 2D axial brain constructed using
the BrainWeb database\cite{aubert-broche_new_2006} with $6,506$
spins. Two variations of the sequence were tested, by rotating the spokes uniformly ($\Delta\theta=\pi/N_{\mathrm{spokes}}$) and by the tiny golden angle ($\Delta\theta=\pi/(\phi+6)$, with $\phi$ the golden ratio).\cite{wundrak_golden_2016}

 The tissue property maps were obtained by performing an external
reconstruction. The methods used were one of the following: 
\begin{itemize}
\item Full-Dict: Filtered back-projection reconstruction for each TR, and then selected the closest dictionary entry by using the maximum dot product.

\item LRTV: Low-rank dictionary matching with Total variation regularization,\cite{golbabaee_compressive_2020,asslander_low_2018,fessler_nonuniform_2003} with a dictionary of a reduced rank of 5.

\end{itemize}

Finally, we compared the quantitative maps on white (WM) and gray
matter (GM) regions by using the Mean Average Error (MAE).

This simulation was run on the same computer as the one used in Section
\ref{subsec:Comparison-with-JEMRIS-Methods}.

\section{Results}

In this section, we report the results obtained from the experiments that compare Koma against other open-source MRI simulators, from the usability tests, and from the MRF showcase.

\subsection{Simulation Accuracy and Speed}

For the simulated examples described in Section~\ref{subsec:Comparison-with-JEMRIS-Methods}, we obtained accurate results with MAEs below 0.1\% when compared to JEMRIS (Figure
\ref{fig:Simulations-(a),-(b),}) and the simulation times were 50, 614,
725, 725, and 1610 times faster, respectively, for Koma. In these tests, Koma improves the simulation
time considerably when the complexity of the problem is increased.

\begin{table}[h]
\caption{\enspace Simulation times for CPU, and GPU acceleration.\label{tab:simulation_times_threads_results}}
\centering{}%
\resizebox{\linewidth}{!}{
\begin{tabular}{cccc}
\toprule 
 & \multicolumn{1}{c}{\textbf{CPU}} & \multicolumn{2}{c}{\textbf{GPU}}\tabularnewline
\midrule 
\textbf{Name } & \textbf{Intel i7-1165G7} & \textbf{GTX 1650 Ti} & \textbf{RTX 2080 Ti}\tabularnewline
\midrule 
JEMRIS & $\approx7\,\mathrm{min}$  & - & -\tabularnewline
MRiLab  & \textbf{1.56~s $\pm$ 0.07~s} & 0.84~s $\pm$ 0.02~s & 0.91~s $\pm$ 0.02~s\tabularnewline
Koma  & 1.82~s $\pm$ 0.17~s & \textbf{0.32~s $\pm$ 0.02~s} & \textbf{0.15~s $\pm$ 0.01~s}\tabularnewline
\bottomrule
\end{tabular}
}
\end{table}

When we tested the simulation speed against MRiLab (Table~\ref{tab:simulation_times_threads_results}), we found that we had slower CPU performance, but we were 2.6 times faster for the GTX 1650Ti and 6.0 times for the RTX 2080 Ti. We think the CPU results show that we still perform unwanted synchronizations between threads, a problem that our GPU implementation would not suffer as we use \texttt{Nthreads=1} by default. An interesting result was that MRiLab was slower for the more powerful RTX 2080 Ti. This is probably explained by CPU-to-GPU memory transfers as the external GPU could be bottle-necked by the Thunderbolt bandwidth capacity. We put most of our attention on the GPU performance, specifically to reduce the number of memory transfers to the GPU by profiling with NVIDIA Nsight tools, which are easily accessed within Julia.

\begin{figure}[h]
\begin{centering}
\includegraphics[width=\linewidth]{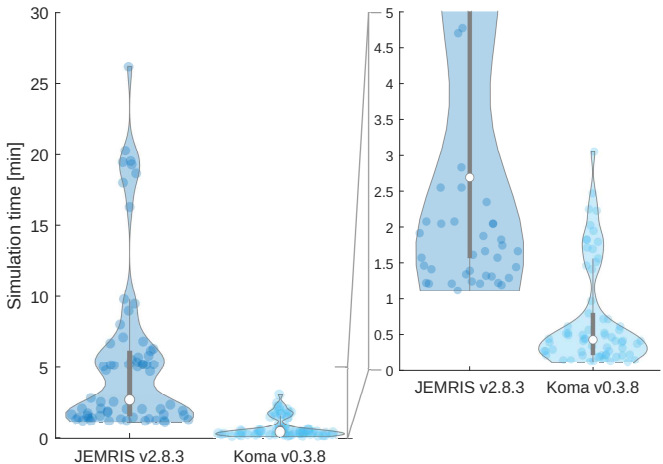}
\par\end{centering}
\caption{\enspace The students' simulation times with their PCs. Each
dot represents one of 6 experiments they needed to make. We only show the results for the 11 students that successfully simulated all the examples.\label{fig:The-students-experience}}
\end{figure}

\subsection{User Experience \label{sub:students_experience_results}}

Students reported no problem installing Julia (mean 4.7/5), Koma (mean 4.2/5), JEMRIS (3.8/5), and MRiLab (4.3/5). Regarding the time taken to install each simulator, most of the students
were able to install Koma (mean 13.2 min), JEMRIS (mean 33.8 min), and MRiLab (mean 16.9 min) in less than 40 minutes.

Their first simulation took them more time in JEMRIS (mean 19 min) and MRiLab (mean 13.9 min) than in Koma (mean 5.7 min). 31\% of the students could not simulate on MRiLab (6 students using Mac OS), so we decided to only use Koma and JEMRIS for the rest of the activities.

To program the pulse sequence, students found that JEMRIS's GUI was slightly better (mean 3.85/5) than Koma's code-based pulse programming (mean 3.69/5). This makes sense since the students' self-reported computational expertise was less than expected (Q1/median/Q3=1.6/2.2/2.6, where 3 meant "I can implement my ideas easily in one programming language"). This feedback helped us improve the pulse sequence programming by implementing our Pulseq file reader, which enables programming the sequence in JEMRIS's GUI. 

Nevertheless, the students also commented that it was unintuitive that
the gradients' strengths in the JEMRIS's GUI were not in $\mathrm{mT/m}$ but scaled by $\gamma$, so $G_{\mathrm{JEMRIS}}=\gamma_{\mathrm{rad/us/mT}}\cdot G_{\mathrm{mT/m}}\approx0.267538 \cdot G_{\mathrm{mT/m}}$. This caused many failed simulations, which prompted us to do an additional tutorial session.

They also readily modify the phantoms with different levels of off-resonance and
motion with JEMRIS (mean 4.27/5) and with Koma (mean 4.24/5).

Finally, their reported median simulation speeds were 8.4 times faster with
Koma than with JEMRIS (Figure~\ref{fig:The-students-experience}), and
65\% ended up recommending Koma over JEMRIS.

\begin{figure*}
\centering{}\includegraphics[width=0.8\textwidth]{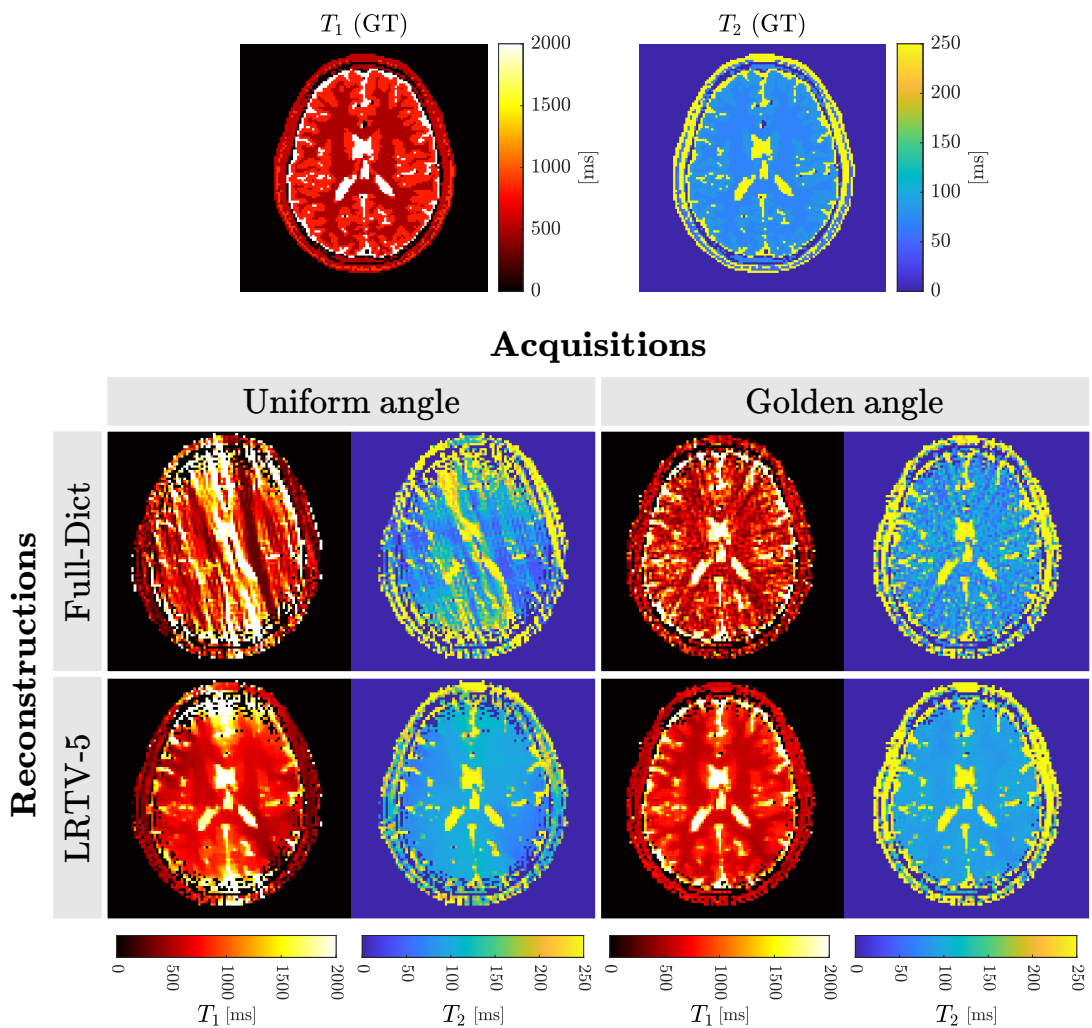}\caption{\enspace Comparison of different MRF acquisitions and reconstruction methods. (GT) Ground-truth,
(Full-Dict) reconstruction using the full dictionary, and (LRTV-5)
Low-rank with Total Variation reconstruction with rank of 5 for the
dictionary. \label{fig:MRF_results}}
\end{figure*}

\subsection{Magnetic Resonance Fingerprinting}

The MRF simulation took approximately 20 seconds to finish. Results
of using different sequences and reconstruction algorithms for our
MRF simulated acquisition are shown in Figure~\ref{fig:MRF_results}.
The best overall pipeline was to rotate the spokes by the tiny golden
angle and then reconstruct using LRTV, which had the lowest MAE's
of [WM-T1, GM-T1, WM-T2, GM-T2]=[55, 26, 15, 13] ms. Thus, we reproduced
results from the state-of-the-art in MRF acquisition just by using
simulations.\cite{wundrak_golden_2016,asslander_low_2018}

This showed the flexibility of our simulator to test novel sequences.
Ideas proven in simulation could potentially be directly executed
on a Pulseq-compatible scanner.\cite{layton_pulseq_2017}

\section{Discussion}

{\bf Efficiency.}
Similarly to other MRI simulators,\cite{puiseux_numerical_2021,xanthis_mrisimul_2014,liu_fast_2017} Koma's variable time-stepping accelerates the simulations, nevertheless, this is not the only contributing factor to efficiency. Another factor is that Koma chooses a different simulation method depending on the sequence regime (excitation or precession). For RF blocks, the method used rotates the magnetization assuming a constant effective field for each time step. For the rest of the sequence (gradient-only or free precession blocks) we assumed a linear effective field for each time step and used trapezoidal integration to estimate the accumulated phase. These methods offer accurate solutions given that we chose the time points carefully, and also that we approximately satisfy the assumptions described in Section~\ref{subsec:Sequence-aware time stepping}, which is generally the case in MRI (even with motion, see Figure~\ref{fig:Simulations-(a),-(b),}). Thus, prior knowledge of the MRI physics and the expected characteristics of $B_1$ and $\boldsymbol{G}(t)$ provides insight to choose a less computationally intensive method needed for an accurate simulation. This was seen in the students experience (\ref{sub:students_experience_results}) where they used a previous version of the software that had uniform time steps. Even then, Koma's MRI-specific solver outperformed the CVODE general ODE solver. This result is not explained by the use of GPUs, as most of the students did not have one in their personal computers. 

In summary, the speed acceleration in our simulator comes from the time-stepping procedure, from the switching between regimes, and from the CPU parallelization and GPU implementation. But there is still room for growth, we can still improve our CPU performance (as shown in Table~\ref{tab:simulation_times_threads_results}), and we can do the same for the GPU implementation. Currently, we are just performing the operations with \texttt{CuArray} types. A proper implementation of some of the functions as GPU kernels, which is also possible in Julia, will potentially further accelerate our simulations. For example, the method to simulate RF-blocks with long soft RF pulses (like adiabatic pulses) is not as efficient as one would expect. The code has already been set up for implementing GPU kernels by using functions that perform in-place operations without returning a result.

{\bf Open Source and Community Standards.}
 Our simulator is open-source, and it is already available on GitHub (\href{https://github.com/cncastillo/KomaMRI.jl}{github.com/cncastillo/KomaMRI.jl}). Furthermore, we seek contributions from the community. For this, we are currently developing documentation with examples using and extending Koma's functionalities. We made and continue to make an effort to make Koma as modular as possible to facilitate its modifications.

We used community-driven and public file formats to increase reproducibility. Koma writes and reads ISMRM raw data, making it compatible with other reconstruction software. Koma's GUI can display and reconstruct raw data acquired on an actual scanner using MRIReco.jl. 

We can also read the newest versions of the Pulseq standard, enabling the generation of the sequence directly in Koma, in MATLAB's Pulseq toolbox, or by using JEMRIS' GUI. If these standards change, the file readers could easily be updated. Our simulator is one of the first to receive Pulseq files as the sequence definition. This will allow us to customize pulse sequences and then test them in real scanners.\cite{loktyushin_mrzero_2021}

The use of MRIReco.jl brings state-of-the-art reconstructions to Koma. This reconstructor can do direct or regularized iterative reconstructions (FISTA,\cite{beck_fast_2009} ADMM,\cite{boyd_distributed_2011} and others). This software also brings flexible and easily customizable reconstructions (see MRIReco.jl documentation). 

 Similarly to MRIReco.jl, we used a dictionary (\texttt{Dict}) to store all the simulation parameters, which can be easily updated to add new parameters. All these parameters are saved to the raw data for later inspection.

{\bf Maintainability and Reproducibility.} 
While performing the experiments for this work, we experienced problems running JEMRIS and MRiLab in modern systems. For JEMRIS, we could not perform multi-threaded simulations out of the box, as some library versions are not supported anymore. We had to compile the package and change the source code to fix some problems with up-to-date versions of CVODE. On the other hand, for MRiLab, the GPU simulations had a similar problem, as it assumed that the system had an old version of CUDA. We had to fix the source code and makefiles to compile for modern versions of CUDA and the MEX libraries. Despite this, we were not able to recover all the functionalities, like the ability to export the signal as ISMRMRD, as it uses a deprecated version of the library.
 
 We believe this experience perfectly represents a common problem for the MR community: maintainability and reproducibility of the software we produce. While not perfect, we believe that Julia helps in minimizing many of these problems. Its modular approach incentives the separation of packages with minimal functionality, which are easier to maintain. Furthermore, all Julia packages are associated with a GitHub page, and to be registered, each new package version is required to maintain or improve the Code Coverage of the tests, and pass the Continuous Integration which assesses the package in multiple versions of Julia and operating systems: Windows, Linux, and Mac. Moreover, even if a package is no longer maintained by the creator, if shared, the \texttt{Manifest.toml} of a package contains all the specific versions of each module, and the environment can be replicated by using the command \texttt{Pkg.activate(".")}, enabling reproducibility. We shared not only the \texttt{Manifest.toml}, but all the code used to replicate the simulations presented in this work in Koma's GitHub.
 
 Julia not only brings high-performance and easy-to-read code but also forces package developers to produce professional software. Decreasing the technical debt passed to new researchers.
 
{\bf Limitations.\label{subsec:Limitations}}
The current implementation of Koma suffers from the same limitations as other Bloch simulators,\cite{liu_fast_2017} which means that some intra-voxel effects, like $T_2^*$ and diffusion, require many spins per voxel, which in turn affects the simulation speed.

An important issue with Bloch simulators is the potential aliasing or spurious echoes when simulating gradient spoilers. They arise due to the finite separation between spins (discrete delta functions) which produces overlapping or aliasing in the Fourier domain. The simplest solution, as before, is to increase the number of spins at the cost of extra computational load. To solve this problem, alternative intermediate solutions should be explored. One of them is to use a different model to describe the spins' state as the one used by the hybrid Bloch-EPG.\cite{guenthner_unifying_2021}

While it is straightforward to implement, we do not yet have multiple coils in our simulator. The lack of coils precludes its ability to simulate conditions at ultra-high fields where coil combination is an issue,\cite{bollmann_challenge_2018} or in highly accelerated sequences where the coil noise characteristics are essential, like in wave-CAIPI.\cite{bilgic_wavecaipi_2015} Also, our simulator is not currently considering some effects, including eddy currents, concomitant gradients, temperature changes, or the drift on the k-space center produced by long readouts.\cite{engel_singleshot_2018}

We designed our simulator to run reasonably fast on a student notebook. More testing is required for more complex scenarios in more powerful servers. Our software has not yet been tested in a multi-GPU system like in Xanthis et al.,\cite{xanthis_high_2014} and more work is needed to take advantage of multiple GPUs.

\section{Conclusions}

In this work, we presented a new general MRI simulator programmed
in Julia. This simulator is fast, easy to use, extensible, open-source, and 
cross-platform. These characteristics were achieved by choosing the appropriate technologies to write easy-to-understand and fast code with a flexible GUI. Furthermore, our simulation method exploits MRI physics and information about the sequence to reduce the simulation times. 

We compared the accuracy of our simulations against JEMRIS, in which we showed high accuracy with MAEs below 0.1\%. We also compared the performance against MRiLab, showing slower CPU times but GPU performance as much as 6 times faster using an RTX 2080 Ti eGPU, and 2.6 times faster using a GTX 1650 Ti.

We also tested the ease of use of Koma with students without previous
knowledge of MRI. Their feedback helped us improve Koma by adding compatibility with community-driven standards like
Pulseq and the ability to load JEMRIS phantoms. Thus, Koma can use the same sequences and phantoms
utilized in JEMRIS. We can also receive simulations from JEMRIS, or scanner-generated raw data, and reconstruct them in our GUI using the exported ISMRMRD file. Moreover, we can export our raw data to the same format and reconstruct the images externally. 

Finally, we showcase the potential to quickly test novel pulse sequences for quantitative MRI before implementing them in the scanner by simulating different MRF acquisitions.

\section{Acknowledgements}

Authors wish to acknowledge funding by the PhD program in Biological and Medical Engineering of the Pontificia Universidad Católica de Chile; by  ANID - Millennium Science Initiative Program ICN2021\_004; by Fondecyt 121074, 1210637, and 1210638; by the Agencia Estatal de Investigaci\'{o}n PID2020-115339RB-I00; by EPSRC EP/P001009, EP/P032311/1, EP/P007619/1, and EP/V044087/1; by Wellcome EPSRC Centre for Medical Engineering NS/A000049/1; ANID Basal FB210024; and by Millennium Nucleus NCN19\_161.

\bibliography{SimulationMRI}

\end{document}